\documentclass[12pt,reqno]{amsart}

\allowdisplaybreaks[4]
\usepackage{graphicx} 
\usepackage{amssymb}
\usepackage{amsmath}
\usepackage{cite}
\usepackage{subfigure}
\usepackage{graphicx}
\usepackage{epstopdf}
\newtheorem{theorem}{Theorem}[section]
\newtheorem{example}[theorem]{Example}

\newtheorem{definition}[theorem]{Definition}

\def\diag{{\rm diag \/ }}

\theoremstyle{plain}
\newtheorem{thm}{Theorem}[section]
\newtheorem{prop}[thm]{Proposition}
\newtheorem{lem}[thm]{Lemma}
\newtheorem{cor}[thm]{Corollary}

\theoremstyle{definition}

\newtheorem{rem}[thm]{Remark}
\newtheorem{defn}[thm]{Definition}
\newtheorem{eg}[thm]{Example}
\newtheorem{subtitle}[thm]{}
\newtheorem{ex}{Exercise}[section]
\numberwithin{equation}{section}

\def\a{\alpha}

\def\l{\lambda}

\def\o{\theta}

\def\cg{{\mathcal{G}}}

\def\cj{{\mathcal{J}}}
\def\ck{{\mathcal{K}}}
\def\cl{{\mathcal{L}}}
\def\cm{{\mathcal{M}}}

\def\cp{{\mathcal{P}}}

\def\cu{{\mathcal{U}}}

\def\p{\partial}

\def\diag{{\rm diag}}

\def\C{\mathbb{C}}

\def\R{\mathbb{R} }
\def\Z{\mathbb{Z}}

\newcommand{\beq}{\begin{equation}}
\newcommand{\eeq}{\end{equation}}
\newcommand{\beg}{\begin{eg}}
\newcommand{\eeg}{\end{eg}}
\newcommand{\bthm}{\begin{thm}}
\newcommand{\ethm}{\end{thm}}
\newcommand{\bprop}{\begin{prop}}
\newcommand{\eprop}{\end{prop}}
\newcommand{\bcor}{\begin{cor}}
\newcommand{\ecor}{\end{cor}}
\newcommand{\blem}{\begin{lem}}
\newcommand{\elem}{\end{lem}}
\newcommand{\bca}{\begin{cases}}
\newcommand{\eca}{\end{cases}}
\newcommand{\brem}{\begin{rem}}
\newcommand{\erem}{\end{rem}}
\newcommand{\bpm}{\begin{pmatrix}}
\newcommand{\epm}{\end{pmatrix}}
\newcommand{\bbm}{\begin{bmatrix}}
\newcommand{\ebm}{\end{bmatrix}}
\newcommand{\bvm}{\begin{vmatrix}}
\newcommand{\evm}{\end{vmatrix}}
\newcommand{\bdefn}{\begin{defn}}
\newcommand{\edefn}{\end{defn}}
\newcommand{\bsub}{\begin{subtitle}}
\newcommand{\esub}{\end{subtitle}}
\newcommand{\bex}{\begin{ex}}
\newcommand{\eex}{\end{ex}}
\newcommand{\ben}{\begin{enumerate}}
\newcommand{\een}{\end{enumerate}}

\begin{document}
\title{Weakly integrable Camassa-Holm-Type equations}
\author{Peilong Dong, Zhiwei Wu, Jingsong He $^{*}$}
\dedicatory{ Department of Mathematics, Ningbo University, Ningbo, 315211 Zhejiang, P.\ R.\ China}

\thanks{*Corresponding author: hejingsong@nbu.edu.cn; jshe@ustc.edu.cn}

\texttt{}
\date{}

\begin{abstract}
Series of deformed Camassa-Holm-Type equations are constructed using the Lagrangian deformation and Loop algebra splittings. They are weakly integrable in the sense of modified Lax pairs.
\end{abstract}

\maketitle

\ \ \ Keywords: Loop algebra splitting, Lagrangian deformation, Camassa-Holm equation\\
\allowdisplaybreaks

\section{Introduction}

The Camassa-Holm (CH) equation  is an important type of shallow water wave equation in fluid dynamics and is known to be integrable \cite{CH93}. It can be written in the following form,
\begin{equation}\label{CH}
u_{t}+2\kappa u_{x}-u_{xxt}+3uu_{x}=2u_{x}u_{xx}+uu_{xxx},
\end{equation}
where $u$ represents the fluid velocity for the $x$ direction and $\kappa$ is the wavenumber. There are many reduction forms of this equation, such as the Benjamin-Bona-Mahony (BBM) equation \cite{BBM72}:
\beq\label{bbm}
u_t+u_x-u_{xxt}+uu_x=0
\eeq
and the Korteweg-de Vries (KdV) equation:
\beq\label{kdv}
u_t=u_{xxx}+6uu_x.
\eeq
Mathematically, \eqref{CH} can be derived from an asymptotic expansion in the Hamiltonian for Euler's equation, or by the tri-Hamiltonian duality from the KdV equation \cite{OR96}.

The CH equation is completely integrable in the sense that it admits a Lax pair, a bi-Hamiltonian structure and  an infinite sequence of conservation laws (cf. \cite{C97,CGI06,L05}). Besides standard traveling wave and (multi) peakon solutions (\cite{ACHM94,CHT04}),  the CH equation has solutions with the presence of breaking waves \cite{CE00}, that is, the wave profile remains bounded while the slope of the wave blows up. This non-local dispersion gains a lot of interests in the research of integrable systems (cf. \cite{C00,CI06,LO00}). Recently, some modifications of the CH equation (CH-type equations) are studied by both mathematicians and physicists. For example, we mention here the modified CH equation with cubic nonlinearity \cite{QLL13} and the $\mu$-CH equation (\cite{KLM08,QFL14}).

Arnaudon \cite{A16a} developed a theory of Lagrangian reduction by using the So\-bo\-lev norm $H^1$ in the Lagrangian instead of the standard $L^2$ norm, from this, deformed equations corresponding to some classical soliton equations can be derived. These deformed equations are called \emph{weakly integrable} in the sense of the deformed Lax pair.

Usually, it is not sure whether the deformed equations are completely integrable. But in some special cases, integrable equations can be obtained. For example, the Camassa-Holm equation can be considered as the deformed equation from the KdV equation. This makes the Lagrangian reduction method highly nontrivial.

Another example is the Camassa-Holm-nonlinear Schr\"{o}dinger (CH-NLS) equation:
\beq\label{cn}
\hat q_t=iq_{xx}\pm2i\hat q(|q|^2-\a^2|q_x|^2), \quad \hat q=q-\a^2 q_{xx}, \a \in \R.
\eeq
Although its integrability is still unknown, its solitary wave solutions preserve some typical soliton properties. Such behavior is studied in Ref. \cite{A16b}. This work motivates us to study the deformed type equations.

One efficient scheme to construct integrable systems in the literature is the loop algebra splitting method; series of soliton equations can be derived from this way (cf. \cite{AKNS74,DS84,SW85,TU99}). For example, the KdV equation, the nonlinear Schr\"{o}dinger equation, the derivative nonlinear Schr\"{o}dinger equation (\cite{F84}) etc. It is well known that when a soliton hierarchy comes from splitting, it is highly possible that we can find a series of B\"{a}cklund or Darboux transformations to construct explicit soltion and rational solutions (cf. \cite{TU00}).

Next we will give a brief introduction of these equations that we will study. All of them describe completely integrable systems. The NLS equation is the simplest and most basic of nonlinear partial differential equations, which has been derived in many branches of physics and admits many types of solutions like breathers, solitons, etc. (cf. \cite{GCGV15,WHXW16}). The bright and dark solitons describe the soliton type pulse propagation in the anomalous and the normal dispersion regimes in optical fibers. We use the freedom in gauge transformation approach to construct dark solitons of the NLS-type equations \cite{VRVP15}. The stationary solutions of NLS equation with periodic condition, can help to present the relationship between four different stationary solutions, then we show how nonlinearity modifies the dynamics of quantum $\delta-$kicked rotator \cite{HLLLN15}. A variety of complex physical systems such as Bose-Einstein condensates \cite{Malomed2016JPB,Panos2016,Bag15,NicolinRRP,ZezyulinRRP,Radha2015,WangRJP2016,Mihalache2014}, nonlinear optical fibers and other types of nonlinear optical waveguides \cite{Leb13,LTM2013,Kolesik2014,Fra14,PetrisPRA,Mihalache2015PRA,Yang2016RJP,LeblondTernicheMihalache}
 etc. usually involve more than one component, so the studies should be extended to multi-component equations, such as the $\mathcal{G}$NLS equation. By using the Darboux transformations, we can derive bright solitons, breathers and rogue waves
 \cite{Ono13,Mihalache2015RRP,Xu2016RRP,Yuan2016,LiuRRP2016,ChenRRP2016,Chen16}  in the self-focusing case, and single dark solitons, multi-dark soliton solutions in the self-defocusing case \cite{LZG13}.

Inspired by the importance of these interesting developments about the analysis of deformed NLS-type equations, we will construct the deformed equations of the derivative NLS (DNLS) equation. The DNLS equation is one of these nonlinear evolution equations, which is of much importance in mathematics and physics. Under special reduction condition, the rogue waves, bright and dark solitons etc., can be obtained by different seed solution \cite{XHW11}. Similarly, the $(2+1)-$dimensional DNLS equation was also investigated \cite{WZ16}. We mention here the Hirota equation, which is a combination of the NLS equation and the modified KdV equation. It is well known that if an integrable equation is a combination of two known equations that are also integrable, the low-order solutions are based on solutions of the two equations \cite{LWWH13}. The same applies to the discrete Hirota equation \cite{GZ16}. And it is decisive to obtain the most general solutions of the Hirota equation, which provide a lot of information about its intrinsic structure, so different types of exact solutions were given in Ref. \cite{BKUI14}.

The main goal of this paper is to combine the Lagrangian deformation under the loop algebra splitting scheme to get a systematic way to construct weakly integrable equations. To illustrate our method, we will show some deformed equations such as the CH-NLS, CH-DNLS, CH-$\cg$NLS, and CH-Hirota equations.

We arrange this paper as follows. In Sec. 2, we introduce the Lie algebra splitting and Lagrangian deformation. In Sec. 3,  we use our method to derive series of deformed soliton equations.  We will end in the last Section with some discussion of the obtained results and future extension of this work.

\section{Lie algebra splitting and Lagrangian deformation}

In this Section, we give a brief review of the loop algebra splitting and Lagrangian deformation.

\noindent \textbf{Loop algebra splitting}.

Let $G$ be a compact Lie subgroup of $GL(n,\C)$, with $\mathcal{G}$ its Lie algebra. Denote $L(G)$ by the group of smooth loops from $S^1$ to $G$, and $\mathcal{L}(\mathcal{G})$ its Lie algebra. That is,
\begin{equation}
\mathcal{L}(\mathcal{G})=\left\{\sum_{i \leq n_0}A_i\lambda^i\mid A_i \in \mathcal{G} ,n_0 \in Z \right\}.
\end{equation}
Let $L(G)_+$ and $L(G)_- $ be subgroups of $ L(G)$, and $L(G)_+\cap L(G)_-=\{e\}$, where $e$ is the identity element. Then we have $ \mathcal{L}(\mathcal{G})=\mathcal{L}(\mathcal{G})_+\oplus \mathcal{L}(\mathcal{G})_-$, which is a direct sum of linear subspaces of $\mathcal{L}(\mathcal{G}) $. A \emph{vacuum sequence} $\cj=\{J_1, J_2, \cdots\}$ is a sequence of commuting elements in $\cl(\cg)_+$. Let $\pi_+$ be a projection of $\cl(\cg)$ onto $\cl(\cg)_+$ along $\cl(\cg)_-$. The \emph{phase space} of evolution is defined as:
\begin{equation}\label{p}
\cm=\pi_+(g_-J_1g_-^{-1}), \quad g_-\in L(G)_-.
\end{equation}
Usually, $\cm=J_1+u$ for some $u \in \cg$. And this is an affine space.

\begin{theorem}\label{ab}(\cite{TU99})
Given $\xi: \R \rightarrow \cm$, there exists a unique $Q_j(\xi) \in \cl(\cg)$ such that:
\begin{equation}\label{lax}
\begin{cases}
[\partial_x+\xi, Q_j(\xi)]=0, \\
Q_j(J_1)=J_j, \quad Q_j(\xi) \text{\ is conjugate to\ } J_j.
\end{cases}
\end{equation}
\end{theorem}
The \emph{$j$-th flow in the G-hierarchy} is of the form:
\begin{equation}
\xi_{t_j}=[\partial_x+\xi, (Q_j(\xi))_+].
\end{equation}

Let $\tau$ be an involution of $G$, such that  $d\tau_{e}$ is conjugate linear in $\mathcal{G}$. Then the fixed point set $\mathcal{U}$  of $\tau$ in $\mathcal{G}$ is a real form. Let $\sigma$ be another complex linear involution on $\mathcal{U}$ and $\mathcal{K(P)}$ the eigenspace of $\sigma$ of eigenvalue$1(-1)$. Hence $\mathcal{U}=\mathcal{K}+\mathcal{P}$ is a Cartan decomposition. For $\xi\in\mathcal{G}$, we use $\xi_\mathcal{G}(\xi_\mathcal{P})$ to denote the $\mathcal{K(P)}$ component of $\xi$. Let

\begin{equation*}
\mathcal{L}_{\tau}(\mathcal{G})=\left\{\sum A(\lambda)\in\mathcal{L}(\mathcal{G})\mid \tau(A(\bar \lambda)) =A(\lambda) \right\}.
\end{equation*}
Consider the following splitting of $\cl_{\tau}(\cg)$:
\begin{equation}
\cl_{\tau}(\cg)_{+}=\{\sum_{i\geq 0}A_{i}\lambda_{i}\in\mathcal{L}(\mathcal{U})\},\quad \cl_{\tau}(\cg)_{-}
=\{\sum_{i\leq 0}A_{i}\lambda_{i}\in\mathcal{L}(\mathcal{U})\}.
\end{equation}

Let $a\in\mathcal{K}$, such that $ad(a^{2})=-id_\cp$. Then given $u\in C^{\infty}(R,\mathcal{P})$, according to Theorem 2.1, there exists unique $Q(u,\lambda)=a\lambda+Q_0(u)+Q_{-1}(u)\lambda^{-1}+\cdots \in \mathcal{L}(\mathcal{U})$ such that $Q(u,\lambda)$ is conjugate to $a\lambda$, and
\begin{equation}\label{ae}
[\partial_x+a\lambda+u, Q(u,\lambda)]=0,
\end{equation}
The \emph{$j$-th flow in the $U$-hierarchy} is
\begin{equation}
[\partial_{x}+a\lambda+u,\partial_{t}+(Q(u,\lambda)\lambda^{(j-1)})_{+}]=0,
\end{equation}
and the Lax pair is the following flat $\mathcal{U}$-value connection one form:

\begin{equation*}
(a\lambda+u)d_{x}+(a\lambda^{j}+u\lambda^{j-1}+\cdots Q_{1-j}(u))dt.
\end{equation*}

\begin{example}[$U(n)$NLS hierarchy]\

Let $\cl(u(n))=\{A(\l)=\sum_iA_i\l^i \mid A_i \in su(n), A(\l)=-\overline{A(\bar{\l})}^t\}$, consider the following splitting of
$\cl(su(n))$:

$$
\begin{cases}
\cl_+(u(n))=\{\sum_{i \geq 0}A_i\l^i \mid A_i \in u(n)\}, \\
\cl_-(u(n))=\{\sum_{i <0} A_i \l^i \mid A_i \in u(n)\}.
\end{cases}
$$
Let $a=\frac{i}{2}I_{k, n-k}$, and $J_1=a\l$, then $\cj=\{a\l^i \mid i \geq 1\}$ is a vacuum sequence. The phase space defined
by \eqref{p} is of the form:

$$
\xi=J_1+u=a\l+\begin{pmatrix} 0 & q \\ -\bar{q}^t & 0\end{pmatrix}, \quad q \in C^{\infty}(\R, \C^{ k\times (n-k)}).
$$
Solve $Q(u, \l)=a\l+Q_0(u)+Q_{-1}(u)\l^{-1}+\cdots \in \cl(u(n))$ from \eqref{lax}, and get

\begin{align*}
& Q_0(u)=u, \quad Q_{-1}(u)=i \begin{pmatrix} -q\bar q^t & q_x \\ \bar{q}^t_x & \bar q^t q \end{pmatrix}, \\
& Q_{-2}(u)=\begin{pmatrix} q_x\bar q^t-q \bar q_x^t & -q_{xx}-2q\bar q^t q \\ \bar{q}^t_{xx}+2\bar q^t q \bar q^t & \bar q_x^tq-\bar q^t q_x\end{pmatrix}.
\end{align*}
The second flow $u_t=[\p_x+u, Q_{-1}]=[Q_{-2}, a]$, which is the matrix NLS equation (or $U(n)$NLS):
\begin{equation}
q_t=i(q_{xx}+2 q\bar q^t q).
\end{equation}
\end{example}
\noindent \textbf{Lagrangian deformation}

\begin{definition}\label{aa}(\cite{A16a})
Let $\mathcal{L}(\mathcal{U})=\sum_{i\leq m_{0}}\ \xi_{i}\lambda^{i}$, $\xi_{i}\in\mathcal{U}$. Define a projection
$P_{k}$ on $\mathcal{L}(\mathcal{U})$ as follows:
\begin{equation}
P_{k}(Z)=\sum_{k<i\leq m_{0}}\Z_{i}\lambda_{i}+(Z_{k-1})_{\mathcal{P}}\lambda_{k-1}+\sum_{i<{k-1}}\Z_{i}\lambda_{i},
\quad Z=\sum_{i\leq m_{0}}\Z_{i}\lambda_{i}.
\end{equation}
\end{definition}
By using the $H^1$-norm instead of $L^2$-norm, the generated operator becomes $\hat \cl=a\l+\hat u$, where $\hat u=u-\a^2u_{xx}$, $\a \in \R$.  Set

$$\hat Q(u)=a\l+u+\hat Q_{-1}(u)\l^{-1}+\hat Q_{-2}(u)\l^{-2}+\cdots.$$

Instead of using \eqref{ae}, we can use the following equation to solve first several coefficients of $\hat Q(u)$, although the process is not purely algebraic.
\beq\label{af}
P_{j}([\p_x+a\l+\hat u, \p_t+(\hat Q(u)\l^{j-1})_+])=0.
\eeq
The \emph{$j$-th deformation flow} is
\beq\label{ag}
(\hat u)_t=((\hat Q_{1-j})_x+[\hat u, \hat Q_{1-j}])_\cp=((\hat Q_{1-j})_\cp)_x+[\hat u, (\hat Q_{1-j})_\ck].
\eeq

Equations constructed through this process are called \emph{weak complete integrable}, and \eqref{af} is called the \emph{weak Lax pair}.

From Lagrangian deformation and the Lax pair of the KdV equation, the CH equation can be derived. In Ref. \cite{A16a}, several deformed equations are derived corresponding to the NLS equation, modified KdV equations and flows constructed from Lie algebra $so(3)$. In the rest of this paper, we will derive some other CH-type equations by the Lie algebra splitting and the weak Lax pairs. For the standard ones (without Lagrangian deformation), more details can be found in Ref. \cite{WH16}.


\section{deformations of integrable hierarchies}

In this Section, we will use the modified loop algebra splitting method with Lagrangian deformation to construct series of new weak completely integrable equations. To explain the algorithm, we carry out in detail the computation of Camassa-Holm-nonlinear Schr\"{o}dinger (CH-NLS) equation as a first example.

\subsection{Camassa Holm-nonlinear Schr\"{o}dinger-type equations}\

In this case, $G=GL(2, \C)$ and $\tau (g)=(\bar g)^{-1}$. Hence the fixed point set of $\tau$ is $\cu=u(2)$. Let $\sigma(g)=I_{1, 1}gI_{1, 1}$, where $I_{1, 1}=\diag(1, -1)$. Then

\begin{align}\label{ac}
& \ck=\R i I_{1, -1}, \quad \cp=\begin{pmatrix} 0 & r \\ -\bar r  & 0\end{pmatrix}.
\end{align}

Let $a=\frac{1}{2}\diag(i, -i)$. Then the phase space $\xi=a\l+u=a\l+\begin{pmatrix} 0 & q \\ -\bar q & 0\end{pmatrix}$ from the standard  loop algebra splitting. And the second flow in the $u(2)$-hierarchy is the NLS equation,
\beq\label{nls}
q_t=i(q_{xx} +2 |q|^2q).
\eeq
By Lagrangian deformation, we have $\hat u=\begin{pmatrix}  0& \hat q \\ -\bar{\hat q} & 0\end{pmatrix}$, where $\hat q=q-\a^2q_{xx}$. Apply \eqref{af} for $j=2$ and compute the coefficient of $\l^i$ in

$$[\p_x+a\l+\hat u, \p_t+(\hat Q(u)\l)_+].$$
\begin{enumerate}
\item[$\l$:] $u_x+[\hat u, u]+[a, \hat Q_{-1}(u)]$, since $u_x \in \cp, [\hat u, u] \in \ck$, $$P_2(u_x+[\hat u, u]+[a, \hat Q_{-1}(u)])=u_x+[a, \hat Q_{-1}(u)]_\cp.$$
From this, we can solve $(\hat Q_{-1}(u))_\cp=i \begin{pmatrix} 0 & q_x \\ \bar q_x & 0 \end{pmatrix}$.
\item[$\l^0$:] $(\hat u)_t=(\hat Q_{-1}(u))_x+[\hat u, \hat Q_{-1}(u)]$.
\\By equating $((\hat u)_t=(\hat Q_{-1}(u))_x+[\hat u, \hat Q_{-1}(u)])_\ck$, we get

\begin{align*}
((\hat Q_{-1}(u))_\ck)_x+[\hat u, (\hat Q_{-1}(u))_\cp]=0.
\end{align*}
Hence $(\hat Q_{-1}(u))_\ck=i(\a^2|q_x|^2-|q|^2)I_{1, 1}$.
\end{enumerate}

Then the $\cp$-component gives the CH-NLS equation \eqref{cn}:

\begin{equation*}
\hat q_t=iq_{xx}-2i\hat q(\a^2|q_x|^2-|q|^2)=iq_{xx}-2i(q-\a^2q_{xx})(\a^2|q_x|^2-|q|^2).
\end{equation*}

\begin{rem} In the process of solving $(\hat Q_{-1}(u))_\ck$, the equation we have is for \\
$((\hat Q_{-1}(u))_\ck)_x$. Therefore, $(\hat Q_{-1}(u))_\ck=i(\a^2|q_x|^2-|q|^2) I_{1, 1}$ up to a constant in $x$. This is different to the standard Lie algebra splitting theory, where this step can be solved algebraically, and it can be proved that this constant is zero.
\end{rem}

\begin{rem} From a direct computation, we can see that $h=|\hat q |^2$ is conserved under the flow \eqref{cn}.
\end{rem}

Next we consider the weak Lax pair for $j=3$. Following a similar computation, we can get

\begin{align*}
&\hat Q_{-1}(u)= i\begin{pmatrix} \a^2|q_x|^2-|q|^2 &  q_x  \\ \bar q_x & |q|^2-\a^2|q_x|^2 \end{pmatrix}, \\
& \hat Q_{-2}(u)=\begin{pmatrix} q_x\bar q-q\bar q_x & -q_{xx}+2(\a^2 |q_x|^2-|q|^2)\hat q \\ \bar q_{xx}-2 (\a^2 |q_x|^2-|q|^2)\bar {\hat q} & q\bar q_x-q_x\bar q\end{pmatrix}.
\end{align*}
Therefore, the third CH-NLS flow (the CH-mKdV equation) is:
\beq\label{cn3}
\hat q_t=-q_{xxx}+2((\a^2 |q_x|^2-|q|^2)\hat q)_x+2(q\bar q_x-q_x\bar q)\hat q.
\eeq


Now we can generalize the argument to $n$-dimensions and derive the CH-$U(n)$NLS equation. Since the process is similar, we only list the result below.

Let $G=GL(n, \C)$ with $\cg=gl(n, \C)$ its Lie algebra. Then $\cu=u(n)$ is the real form under the involution $\tau(g)={\bar g}^t$. Let $\sigma$ be the involution of $\cg$ defined by $\sigma(g)=I_{k, n-k}gI_{k, n-k}$, where $I_{k, n-k}=\diag(I_k, -I_{n-k})$.
Then
\begin{align*}
\ck= u(k) \times u(n-k), \quad \cp=\begin{pmatrix} 0 & X \\ -\bar X^t & 0 \end{pmatrix}, X \in \C^{k \times (n-k)}.
\end{align*}
Let $a=\frac{i}{2}I_{k, n-k}$, then $u=\begin{pmatrix} 0 & q \\ -\bar q^t & 0\end{pmatrix}$, where $q \in \C^{k \times (n-k)}$. Under Lagrangian decomposition, we get  $\hat{u}=\begin{pmatrix} 0 & \hat{q} \\ -\bar {\hat{q}}^t & 0\end{pmatrix}$, with  $\hat q= q-\a^2 q_{xx}$. From the weak Lax pair \eqref{af} for $j=2$, we get

$$\hat Q_{-1}(u)= i \begin{pmatrix} \a^2q_x\bar q_x^t-q\bar q^t & q_x \\ \bar q^t_x& \bar q^t q-\a^2\bar q_x^tq_x\end{pmatrix}.$$
Therefore, the CH-$U(n)$NLS equation is:
\beq
\hat q_t=iq_{xx}+i(\hat q(\bar q^t q-\a^2\bar q^t_x q_x)+(q\bar q^t-\a^2q_x\bar q^t_x)\hat q).
\eeq
In particular, when $n=2$, this equation becomes \eqref{cn}.

\subsection{The CH-DNLS equation}\

In this Section, we start with the Lax pair of the DNLS equation. Then we consider the constraint under the Lagrangian deformation. The construction of the DNLS equation can be found in Ref. \cite{WH16}.

Let $G=SU(2)$, and $L(SU(2))$ be the group of smooth loops from $S^1$ to $SU(2)$, and $\cl(su(2))$ its Lie algebra.
Define an involution $\sigma$ on $su(2)$ as following:
\beq\label{s}
\sigma (A)=I_{1, 1}AI_{1, 1}^{-1}, \quad I_{1,1}=\diag(1, -1).
\eeq
Let $\ck$ and $\cp$ denote the $1$ and $-1$ eigenspaces of $\sigma$, respectively, then

$$\ck=\R i I_{1, 1}, \quad \cp=\begin{pmatrix} 0 & r \\ -\bar r & 0 \end{pmatrix}. $$
Furthermore, $\sigma$ induces an involution on $\cl(sl(2, \C))$ such that

$$\sigma (A(\l))=I_{1,1}A(-\l)I_{1, 1}^{-1}.$$
Let $\cl_{\sigma}(su(2))$ be the subalgebra of $\cl(su(2))$ consisting of fixed points of $\sigma$, and consider the following splitting of $\cl(su(2))$:

\begin{align*}
\begin{cases}
\cl_\sigma^+(su(2))=\{\sum_{i \geq 1}A_i \l^i \in \cl_\sigma(su(2))\}, \\\cl_\sigma^-(su(2))=\{\sum_{i \leq 0} A_i \l^i \in \cl_\sigma(su(2))\}.
\end{cases}
\end{align*}
Given $u=\begin{pmatrix} 0 & q \\ -\bar { q}^t & 0 \end{pmatrix}$, write

$$Q(u, \l)=a\l^2+Q_1(u)\l+Q_0(u)+Q_{-1}(u)\l^{-1}+\ldots \in \cl_\sigma(su(2)).$$
Such $Q(u, \l)$ can be solved uniquely by the following system:
\begin{equation*}
\begin{cases}
[\p_x+a\l^2+u\l, Q(u, \l)]=0, \\
Q(u, \l)^2=-\l^4 I_2.
\end{cases}
\end{equation*}
Then the DNLS equation can be written as
\beq\label{dnls}
q_t=iq_{xx}-(|q|^2q)_x.
\eeq

Now we construct the deformed equation for $\hat u=\begin{pmatrix} 0 & \hat q \\ -\bar {\hat q}^t & 0 \end{pmatrix}$ as follows.
Write $\hat Q(u, \l)=a\l^2+u\l+\hat Q_0(u)+\hat Q_{-1}(u)\l^{-1}+\cdots$, with

$$\hat Q_0(u)=\begin{pmatrix} A & 0 \\ 0 & -A\end{pmatrix}, \quad \hat Q_{-1}(u)=\begin{pmatrix} 0& B \\ -\bar B & 0\end{pmatrix}.$$
Consider the following weak Lax pair:
\beq\label{ai}
P_{5}([\p_x+a\l^2+\hat u\l, \p_t+(\hat Q(u)\l^{2})_+])=0,
\eeq
and compare the coefficient of $\l^j$ of \eqref{ai}. The first non-trivial equation comes from the coefficient of $\l^3$:

$$u_x+[a, \hat Q_{-1}(u)]+[\hat u, \hat Q_{0}(u)]=0.$$
Therefore,
\beq\label{aj}
q_x+iB-2A\hat q=0.
\eeq
Compute the coefficient of $\l^2$ in \eqref{ai} to get:

$$(\hat Q_0(u))_x+[\hat u, \hat Q_{-1}(u)]=0.$$
Hence $A_x=\hat q \bar B-B\bar{\hat q}$. Together with \eqref{aj} and note that $\bar A=-A$, we can solve $A=i(\a^2|q_x|^2-|q|^2)$ (up to a constant in $x$).

Therefore, the CH-DNLS equation is
\beq\label{ak}
\hat q_t=iq_{xx}+2((\a^2|q_x|^2-|q|^2)\hat q)_x, \quad \hat q=q-\a^2q_{xx}.
\eeq

\begin{rem} In Ref. \cite{WH16}, the authors used the loop algebra splitting method to get hierarchies of the generalized DNLS equations, for example,
\beq\label{dn}
q_t=\frac{i}{2}q_{xx}-(2\o+1)|q|^2q_x-(2\o-\frac{1}{2})q^2\bar{q}_x+(\frac{1}{2}\o+2\o^2)i|q|^4q, \quad \o \in \R.
\eeq
By choosing different values of $\o$, the DNLSI, DNLSII, and DNLSIII are derived.

But the deformed DNLSII and DNLSIII equations do not come from the weak Lax pair. That is, equation in the form of \eqref{ai} is not solvable. This means we may need to find more reductions in order to make it well-posed.
\end{rem}

\subsection{CH-Hirota equation}\

The Hirota equation is a combination of the NLS equation and the mKdV equation \cite{H73}:
\beq\label{h}
q_t=\beta i (q_{xx}+2q|q|^2)-\gamma(q_{xxx}+6|q|^2q_x), \quad  \beta, \gamma \in \R.
\eeq
It can be derived from the following Lax pair:

\begin{equation*}
[\p_x+a\l+u, \p_t+\gamma a \l^3 + (\beta a +\gamma u)\l^2+Q_{-1}(u)\l+Q_{-2}(u)]=0,
\end{equation*}
where

\begin{align*}
& a=\frac{i}{2} \begin{pmatrix} 1 & 0 \\ 0 & -1 \end{pmatrix}, \quad u=\begin{pmatrix} 0 & q \\ -\bar q & 0\end{pmatrix}, \quad  Q_{-1}(u)= \begin{pmatrix}-\gamma i |q|^2 & \gamma i q_x+\beta q \\
\gamma i \bar q_x-\beta \bar q & \gamma i |q|^2  \end{pmatrix}, \\
& Q_{-2}(u)= \begin{pmatrix} -\gamma (q \bar q_x- q_x\bar q)-\beta i |q|^2 &  \beta i q_x-\gamma q_{xx}-2 \gamma q |q|^2 \\
\beta i \bar q_x+\gamma \bar q_{xx} + 2 \gamma \bar q |q|^2 & \gamma (q \bar q_x- q_x\bar q)+\beta i |q|^2\end{pmatrix}.
\end{align*}

Under the Lagrangian reduction,  where $G=SU(2)$, $\ck$, $\cp$ are as in \eqref{ac},  and $\hat{u}=\left(\begin{matrix} 0&\hat{q} \\
-\hat{\bar{q}}&0\end{matrix}\right), \hat{q}=q-\a^{2}q_{xx}$.

Solve $\hat{Q}(u,\lambda)=\hat{Q}(u,\lambda)=4\gamma a\lambda+(2\beta a+4\gamma u)+\hat{Q}_{-1}\lambda^{-1}+\ldots \in \cl(su(2))$ from
\begin{equation}
P_{3}([\partial_{x}+a\lambda+
\hat{u},\partial_{t}+(\hat{Q}(u)\lambda^{2})_{+}])=0,
\end{equation}
where $P_3$ is the projection defined in order to get

\begin{align*}
& \hat Q_{-1}(u)=\begin{pmatrix} -\gamma i(|q|^2-\a^2|q_x|^2) & \beta \hat q +\gamma i q_x \\ -\beta \bar {\hat q} + \gamma i {\bar q}_x & \gamma i(|q|^2-\a^2|q_x|^2) \end{pmatrix},  \\
& \hat Q_{-2}(u)=\begin{pmatrix} \hat Q_{-2, 11}(u) &\hat Q_{-2, 12}(u) \\
-\overline{\hat Q_{-2, 12}(u)} &-\hat Q_{-2, 11}(u) \end{pmatrix},
\end{align*}
where

\begin{align*}
&\hat Q_{-2, 11}(u)=-i\beta |\hat q|^2-\gamma (q {\bar q}_x-q_x\bar q),  \\
&\hat Q_{-2, 12}(u)=\beta i \hat q_x- \gamma q_{xx}-2\gamma \hat q (|q|^2-\a^2 |q_x|^2).
\end{align*}
Then the CH-Hirota equation is
\beq
\hat q_t=\beta i \hat q_{xx}- \gamma q_{xxx}-2\gamma (\hat q (|q|^2-\a^2 |q_x|^2))_x+2\hat q (i\beta |\hat q|^2+\gamma (q {\bar q}_x-q_x\bar q)).
\eeq

\section{Conclusion}
In this paper, we tried to understand the Lagrangian deformation theory in Ref. \cite{A16a} in terms of loop algebra splitting. By adding the Lagrangian reduction in the hierarchies constructed by using the loop algebra splitting, such as the $U(n)$NLS hierarchy, the DNLS equation, and the Hirota equation, we get a series of deformed equations.

This give us an algorithm to construct new weakly integrable equations. In particular, by choosing different Lie algebra $\cg$ and real form $\cu$, from the Cartan decomposition and splitting of the loop algebra $\cl(\cu)$, we can get the corresponding deformed equations. As pointed out in Sec. 3, the weak Lax pair depends on the choice of the projection $P_j$ as defined in Definition \ref{aa}. And it can be checked that it may not work for the whole series, for example, in the case of the CH-NLS equation, when we consider the deformed equations corresponding to the forth or higher flows in the hierarchy, we may not get a well-posed equation. The next issue of interest would be find a modification that works for the whole hierarchy.

Note that the ``integrability" of these equations is still an open problem: so far, we only know that the CH equation is completely integrable. Therefore, it would be worthwhile to find soliton-like solutions for these equations, which may give us a better understanding of these new types of equations. Work on this direction is underway.
Also, it is well known that if an equation is completely integrable, then there exists a family of infinitely many conservation laws. As we noted in Sec. 3, another future research direction is to write down the conserved quantities for the weakly integrable equations.

\emph{acknowledgement}
This research was supported in part by the NSF of China under Grants Nos. 11401327 and 11671219, by the K.C. Wong Magna Fund in Ningbo University and the Scientific Research Foundation of Graduate School of Ningbo University.

\end{document}